\documentclass[preprint,showpacs,preprintnumbers,amsmath,amssymb, superscriptaddress]{revtex4}

\usepackage{textcomp}
\usepackage{makeidx}
\usepackage{amsmath}
\usepackage{subfigure}
\usepackage{amssymb}
\usepackage{hyperref}
\usepackage{graphicx}
 \usepackage{epstopdf}
\usepackage[utf8]{inputenc}
\usepackage{float}
\usepackage{color}

\begin{document}

\title{A new family of analytical anisotropic solutions by gravitational decoupling}

\author{Milko Estrada}
\email{mi.estrada@profesor.duoc.cl}
\affiliation{Departamento de F\'isica, Facultad de ciencias básicas, Universidad de Antofagasta, Casilla 170, Antofagasta, Chile.}
\affiliation{Instituto de Matem\'atica, F\'isica y Estad\'istica, Universidad de las Am\'ericas,
Manuel Montt 948, Providencia,Santiago, Chile.}
\author{Francisco Tello-Ortiz}
\email{ftelloortiz@gmail.com}
\affiliation{Departamento de F\'isica, Facultad de ciencias básicas, Universidad de Antofagasta, Casilla 170, Antofagasta, Chile.}

\date{\today}

\begin{abstract}
This work is focused in the study of analytic anisotropic solutions to Einstein's field equations, describing spherically symmetric and static configurations by way of the gravitational decoupling through the method of Minimal Geometric Deformation (MGD). For this we apply MGD to Heintzmann's solution obtaining two new analytic and well behaved anisotropic solutions, in which all their parameters such as the effective density, the effective radial and tangential pressure, as well as radial and tangential sound speed, fulfill each of the requirements for the physical acceptability available in the literature.
 
\end{abstract}


\maketitle

\section{Introduction}
In recent years has emerged a great interest in the anisotropic solutions of Einstein's equations. This is so because the  anisotropic scenarios describe some relevant astrophysical phenomena. Harko and Mak \cite{Harko2} showed some examples of this: nuclear matter may be anisotropic in certain high density ranges, or from the point of view of the Newtonian gravity, spherical galaxies can have anisotropic matter distribution. Harko and Mak in the references \cite{Harko2,Harko1} argue that the interior of a star must fulfill the general physical conditions that describe a well behaved isotropic or anisotropic solution . Due to the later arguments, several models of anisotropic compact objects and its criteria have been studied in the literature. \cite{anisotropico1,anisotropico2,anisotropico3,anisotropico4,anisotropico5,anisotropico51,anisotropico52}.

According to the above mentioned and other arguments,  in recent years there has risen a great interest in finding new anisotropic analytic solutions of Einstein equations, but this is not an easy task due the highly nonlinear behaviour of these equations (some examples of how to generate anisotropic solutions are exposed at references \cite{anisotropico6,anisotropico7,anisotropico8}). To address this problem,  J. Ovalle \cite{Ovalle1}, proposed a method called {\it Minimal Geometric Deformation} (MGD), which leads to finding new analytic anisotropic solutions of Einstein equations for spherically symmetric and static configurations. MGD was initially employed for the study of solutions at the braneworld  scenarios \cite{Ovalle3,Ovalle4} and then this was spread to the study of black hole solutions  \cite{Ovalle5,Ovalle6} (other applications can be seen on references \cite{Ovalle7,Ovalle8,Ovalle9,Ovalle10}).

In this method the isotropic energy momentum tensor  $\bar{T}_{\mu\nu}$  is deformed by an additional source $\Theta_{\mu\nu}$ whose coupling is proportional to the constant $\alpha$, and causes anisotropic effects on the self-gravitating system. This additional source can contain new fields, like scalar, vector and tensor fields \cite{Ovalle2}. However in this work the source  $\Theta_{\mu \nu}$ will represents a generic gravitational source. Then the energy momentum tensor reads:
\begin{equation}
T_{\mu\nu} \mapsto \bar{T}_{\mu \nu} + \alpha \Theta_{\mu\nu}, \label{desacoplamiento}
\end{equation}
with the corresponding conservation equation:
\begin{equation}
    \nabla_\nu T^{\mu \nu} =0. \label{conservacion}
\end{equation}

Ovalle at reference \cite{Ovalle2} argue ``To summarise, the MGD-decoupling amounts to the following procedure: given two gravitational
sources A and B, standard Einstein’s equations are first solved for A, and then a simpler
set of quasi-Einstein equations are solved for B. Finally, the two solutions can be combined in
order to derive the complete solution for the total system ''. The remarkable of this method is that: one isotropic solution A is deformed and it produces a combined anisotropic solution $A \cup B$ that preserves spherical symmetry. Due to the above explanation it is very interesting to apply MGD to an isotropic well behaved and spherically symmetric solution and then analyze the behavior of the new anisotropic solution (but now with an anisotropic criterion). Regarding this, in reference 
\cite{Lake} one hundred twenty seven isotropic solutions were analyzed where only $9$ of them were well behaved from physical point of view.

One interesting isotropic spherically symmetric solution is the Heintzmann's space time \cite{Heint}. In reference \cite{Lake} it was shown that this solution is well behaved for an arbitrary election of the constants (specifically the special case in which the constants $a$, $c$ and $A$ have magnitude equal to 1 on equation (\ref{heint})). In this work we will show that this solution is still well behaved for values of mass and radii of physical interest described below. These values yield to values of constants $a$ and $c$ that are no longer restricted to be equal to one. Heintzmann's solution has been extended to an anisotropic charged case in the references \cite{Heint1,Heint2} also showing that it is a well behaved solution.

Furthermore, in this work we will analyze the minimal geometric deformation of isotropic Heintzmann's solution and we will get two new anisotropic solutions. The matching conditions are obtained with the Schwarzschild exterior solution and then we studied the physical admissibility of these new solutions. For this, we will use realistic values of radii and mass that correspond to the starts 4U 1538-52, RXJ 1856-37 and Vela X-1  \cite{anisotropico51,anisotropico52}, again these values of mass and radii will lead to values of the constants $a$ and $c$  that are no longer restricted to be equal to one.

This work is organized as follows: section II presents the Einstein equations for the energy momentum tensor (\ref{desacoplamiento}), in section III we explain in more details the MGD method, in section IV we study the physical acceptability of the isotropic Heintzmann's solution for typical values of radius and mass of some compact objects. Section V is devoted to the application of the MGD method to the Heintzmann's solution, showing two new anisotropic physically acceptable solutions, finally section VI summarize the essentials of this work and exposes some conclusions. 

\section{Einstein's field equations for multiple sources}

Starting with the standard Einstein's equations:

\begin{equation}\label{einsteineq1}
    G_{\mu \nu}\equiv R_{\mu\nu}-\frac{1}{2}Rg_{\mu\nu}=-\kappa^2 T_{\mu \nu}, 
\end{equation}
where $T_{\mu \nu}$ is given by (\ref{desacoplamiento}) and $\bar{T}_{\mu \nu}$ corresponds to a perfect fluid: 

\begin{equation}\label{perfect}
\bar{T}_{\mu \nu }=(\bar{\rho} +\bar{p})\,u_{\mu }\,u_{\nu }-\bar{p}\,g_{\mu \nu },
\end{equation}
being $u^\mu$ the four-velocity, $\bar{\rho}$ and $\bar{p}$ the density and the isotropic pressure respectively.

In Schwarzschild coordinates the spherically symmetric line element reads:

\begin{equation}\label{metric}
ds^{2}=e^{\nu (r)}\,dt^{2}-e^{\lambda (r)}\,dr^{2}-r^{2}\left( d\theta^{2}+\sin ^{2}\theta \,d\phi ^{2}\right),
\end{equation}
where $\nu=\nu(r)$ and $\lambda=\lambda(r)$ are purely radial functions and $r$ ranging from $r=0$ (the object center) to $r=R$ (the object surface). The line element (\ref{metric}) satisfy the Einstein's equations (\ref{einsteineq1}), which leads to:
\begin{eqnarray}\label{tt}
\kappa^2
\left(
\bar{\rho}+\alpha\,\Theta_0^{0}
\right)
&\!\!=\!\!&
\frac 1{r^2}
-
e^{-\lambda }\left( \frac1{r^2}-\frac{\lambda'}r\right)\ ,
\\
\label{rr}
\kappa^2
\left(\bar{p}-\alpha\,\Theta_1^{1}\right)
&\!\!=\!\!&
-\frac 1{r^2}
+
e^{-\lambda }\left( \frac 1{r^2}+\frac{\nu'}r\right)\ ,
\\
\label{fifi}
\kappa^2
\left(\bar{p}-\alpha\,\Theta_2^{2}\right)
&\!\!=\!\!&
\frac {e^{-\lambda }}{4}
\left( 2 \nu''+\nu'^2-\lambda' \nu'
+2 \frac{\nu'-\lambda'}r\right),
\end{eqnarray}

The conservation equation (\ref{conservacion}), which is a linear combination of Eqs. (\ref{tt})-(\ref{fifi}), yields:

\begin{equation} \label{conservacionA}
    \bar{p}'+\frac{\nu'}{2} (\bar{\rho}+\bar{p})- \alpha (\Theta^1_1)'+\frac{\nu'}{2}\alpha (\Theta^0_0-\Theta^1_1)+\frac{2\alpha}{r}(\Theta^2_2-\Theta^1_1)=0.
\end{equation}

Here the prime means differentiation respect to $r$. So, the perfect fluid is recovered in the limit $\alpha \to 0$. 

In order to simplify the above system, we identify the effective density, the effective radial and tangential pressures as:
\begin{eqnarray}
    \rho=\bar{\rho}+ \alpha  \Theta^0_0 \label{rhoeff} \\
    p_r=\bar{p}-\alpha  \Theta^1_1 \label{pradialeff}\\
    p_t=\bar{p}-\alpha  \Theta^2_2, \label{ptangencialeff}
\end{eqnarray}
where in the extra fluid $\Theta^1_1 \neq \Theta^2_2 =\Theta^3_3$. So the total anisotropy introduced by the generic source $\Theta_{\mu\nu}$ is given by:
\begin{equation}\label{anisotropy}
\Pi
\equiv
p_t-p_r
=
\alpha\left(\Theta_1^{1}-\Theta_2^{2}\right)
\ 
\end{equation}

\section{Minimal Geometric Deformation Method}

This method for spherically symmetric distributions was proposed in references \cite{Ovalle1,Ovalle2}. We will begin by considering a solution to Eqs. (\ref{tt})-(\ref{fifi}) with $\alpha= 0$, namely, a GR perfect fluid solution $\{\eta, \mu, \bar{\rho}, \bar{p} \}$, where $\eta$ and $\mu$ are the corresponding metric functions:

\begin{equation}
ds^2=e^{\eta(r)}dt^2-\mu(r)^{-1}dr^2-r^2 d\Omega^2. \label{elementodelinea1}
\end{equation}

Turning on the parameter $\alpha$ we can see the effects of the
source $\Theta_{\mu\nu}$ on the perfect fluid solution $\{\eta,\mu\,\bar{\rho},\bar{p}\}$. These effects can be encoded in the geometric deformation undergone by the perfect fluid geometry $\{\eta,\mu\}$ in equation (\ref{elementodelinea1}) as follows:
\begin{equation}\label{mu}
\mu(r)\mapsto e^{-\lambda(r)}=\mu(r)+\alpha f^{*}(r)    
\end{equation}
\begin{equation}\label{nu}
\eta(r)\mapsto \nu(r)=\eta(r),    
\end{equation}
it means that only the radial component of the line element  (\ref{elementodelinea1}) is deformed, where $f^{*}(r)$ is the corresponding deformation to the radial part. Upon replacing equations (\ref{mu}) and (\ref{nu})  in the Einstein equations (\ref{tt})-(\ref{fifi}), the system splits into two sets of equations :

\begin{enumerate}
    \item The standard Einstein equations for a perfect fluid (with $\alpha=0$), where $\eta(r)=\nu(r)$:

\begin{eqnarray}\label{ro}
\kappa^{2} \bar{\rho}&=&\frac{1}{r^{2}}-\frac{\mu}{r^{2}}-\frac{\mu^{\prime}}{r}\\\label{pr}
\kappa^{2} \bar{p}&=&-\frac{1}{r^{2}}+\mu\left(\frac{1}{r^{2}}+\frac{\nu^{\prime}}{r}\right)\\\label{pt}
\kappa^{2} \bar{p}&=&\frac{\mu}{4}\left(2\nu^{\prime\prime}+\nu^{\prime2}+2\frac{\nu^{\prime}}{r}\right)+\frac{\mu^{\prime}}{4}\left(\nu^{\prime}+\frac{2}{r}\right),
\end{eqnarray}
along with the conservation equation (\ref{conservacion}) with $\alpha=0$, namely $\nabla_{\nu}\bar{T}^{\mu\nu}$ , yielding:
\begin{equation}\label{conservde}
\bar{p}^{\prime}+\frac{\nu^{\prime}}{2}\left(\bar{\rho} +\bar{p} \right)=0,    
\end{equation}
which is a linear combination of equations (\ref{ro})-(\ref{pt}).

\item 
The terms of order $\alpha$ give rise to the following {\it quasi-Einstein equations} \cite{Ovalle1,Ovalle2} , which includes the source $\Theta_{\mu \nu}$:

\begin{eqnarray}\label{cero}
\kappa^{2}\Theta^{0}_{0}&=&-\frac{f^{*}}{r^{2}}-\frac{f^{*\prime}}{r} \label{uno} \\ 
\kappa^{2}\Theta^{1}_{1}&=&-f^{*}\left(\frac{1}{r^{2}}+\frac{\nu^{\prime}}{r}\right) \label{dos} \\ 
\kappa^{2}\Theta^{2}_{2}&=&-\frac{f^{*}}{4}\left(2\nu^{\prime\prime}+\nu^{\prime2}+2\frac{\nu^{\prime}}{r}\right)-\frac{f^{*\prime}}{4}\left(\nu^{\prime}+\frac{2}{r}\right). \label{tres}
\end{eqnarray}

The conservation equation (\ref{conservacion}) then yields to $\nabla_{\nu}\Theta^{\mu\nu}=0$, which explicitly reads:
\begin{equation} \label{conservacionB}
\left(\Theta^{1}_{1}\right)^{\prime}-\frac{\nu^{\prime}}{2}\left(\Theta^{0}_{0}-\Theta^{1}_{1}\right)-\frac{2}{r}\left(\Theta^{2}_{2}-\Theta^{1}_{1}\right)=0.  
\end{equation}
\end{enumerate}

It is worth to stress that equations (\ref{conservde}) and (\ref{conservacionB}) imply that there is no exchange of energy momentum between the perfect fluid and the extra source $\Theta^\mu_\nu$. So only there is purely gravitational interaction.

\section{Heintzmann's solution} \label{isotropia}

In this section we will analyze if the isotropic Heintzmann's solution \cite{Heint} is still well behaved for our physical interest parameters of mass and radii, these correspond to the starts 4U 1538-52, RXJ 1856-37 and Vela X-1 \cite{anisotropico51,anisotropico52}. In reference \cite{Lake} this solution is called as Heint IIa, The line element is: 

\begin{equation}\label{heint}
ds^{2}=A^{2}\left(1+ar^{2}\right)^{3}dt^{2}-\left(1-\frac{3ar^{2}}{2}\frac{1+c\left(1+4ar^{2}\right)^{-1/2}}{1+ar^{2}}\right)^{-1}dr^{2}-r^{2}d\Omega^{2},    
\end{equation}
where $a$, $c$ and $A$ are constant parameters. The pressure $\bar{p}$  and the energy density $\bar{\rho}$ are:
\begin{eqnarray}\label{presionheint}
\bar{p}(r)&=&-\frac{3a\left[\left(3ar^{2}-3\right)\left(1+4ar^{2}\right)^{1/2}+c\left(1+7ar^{2}\right)\right]}{2\kappa^{2}\left(4ar^{2}+1\right)^{1/2}\left(1+ar^{2}\right)^{2}} \\ \label{densidadheint}
\bar{\rho}(r)&=&\frac{3a\left[\left(4a^{2}r^{4}+13ar^{2}+3\right)\left(1+4ar^{2}\right)^{1/2}+3c+9acr^{2}\right]}{2\kappa^{2}\left(4ar^{2}+1\right)^{1/2}\left(1+ar^{2}\right)^{2}}. \label{rho}
\end{eqnarray}

\subsection{Coupling with Schwarzschild exterior solution}

We will present the conditions for the coupling of isotropic solution (\ref{heint}) with the Schwarzschild vacuum exterior solution:

\begin{equation}\label{Schwarzschilds}
ds^{2}=\left(1-\frac{2\bar{M}}{r}\right)dt^{2}-\left(1-\frac{2\bar{M}}{r}\right)^{-1}dr^{2}-r^{2}d\Omega^{2}.    
\end{equation}

For this, we will take into account the Israel-Darmois matching conditions. At the stellar surface $\Sigma$ defined by $r=R$ these conditions give \cite{Israel}:

\begin{equation}\label{secondform}
    \left[{G}_{\mu\nu}r^{\nu}\right]_{\Sigma}=0, 
\end{equation}
where $r_{\nu}$ is an unit radial vector and  $\left[F\right]_{\Sigma}\equiv F\left(r \rightarrow R^{+}\right)- F\left(r \rightarrow R^{-}\right)$, for any function $F=F(r)$. Using equation (\ref{secondform}) and the general Einstein equations, we find that:
\begin{equation}
    \left[\bar{T}_{\mu\nu}r^{\nu}\right]_{\Sigma}=0,
\end{equation}
which leads to:
\begin{equation}\label{secondformpr}
    \left[\bar{p}\right]_{\Sigma}=0,
\end{equation}
because in the external vacuum Schwarzchild solution the pressure is zero, then the last equation yields:
\begin{equation}
\bar{p}(R)=0.    \label{p4}
\end{equation}
Using the equation (\ref{presionheint}), we find the value of the constant $c$ from equation (\ref{p4}):
\begin{equation}
c=\frac{3\left(1+4aR^{2}\right)^{1/2}\left(1-aR^{2}\right)}{1+7aR^{2}} \label{c1}.
\end{equation} 

The second condition, also given by the Israel-Darmois conditions, says that the line element must be continuous {\it i.e.} there must be no jumps in the metric:

\begin{equation}
    \left[ds^{2}\right]_{\Sigma}=0, \label{firstform}
\end{equation}
so,

\begin{equation}\label{firstnu}
    g_{tt}^{\,\,-}(R)=g_{tt}^{\,\,+}(R),
\end{equation}
and:
\begin{equation}\label{firstlambda}
    g_{rr}^{\,\,-}(R)=g_{rr}^{\,\,+}(R),
\end{equation}
where $g_{tt}$ and $g_{rr}$ are the temporal and radial components of the metric. Then equating (\ref{heint}) with (\ref{Schwarzschilds}) and using the conditions (\ref{firstnu}) and (\ref{firstlambda}) we get:

\begin{equation}
A^{2}\left(1+aR^{2}\right)^{3}=1-\frac{2\bar{M}}{R},    \label{continuidadheintt}
\end{equation}
\begin{equation}
  1-\frac{3aR^{2}}{2}\frac{1+c\left(1+4aR^{2}\right)^{-1/2}}{1+aR^{2}}=1-\frac{2\bar{M}}{R}. \label{continuidadheintr}
\end{equation}

We will take $\bar{M}$ and $R$ as free parameters. Inserting the parameter $c$ from equation (\ref{c1})  into the equation (\ref{continuidadheintr}), we find:

\begin{equation}
   a=\frac{\bar{M}}{R^2(3R-7\bar{M})}, \label{a1}
\end{equation}
then with this value for $a$ we obtain the value for $c$ from equation (\ref{c1}) and for $A$ from equation (\ref{continuidadheintt}).

\begin{figure}
\centering 
\includegraphics[width=7.5in]{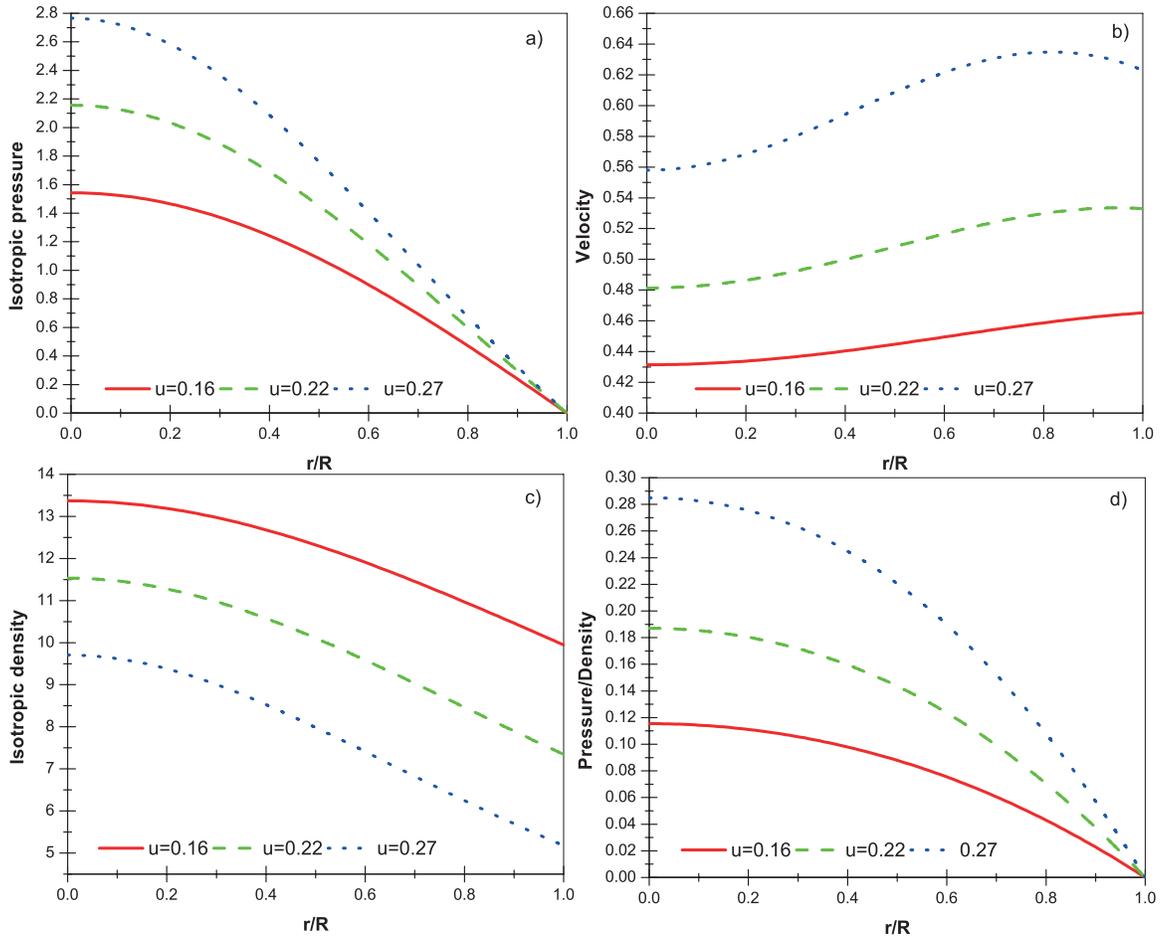}
\caption{The panel $a)$ shows the isotropic pressure which vanishes at the boundary $r=R$. Panel $b)$ shows that the sound speed velocity obeys  causality condition $v_{s}\leq 1$. Panel $c)$
exhibits the monotonically decreasing behaviour of the density with increasing $r$. Finally panel $d)$ shows the decreasing pressure-density ratio $\bar{p}/\bar{\rho}$. In the graphics the pressure and density are divided by the value of $a$. These plots correspond to solution of section \ref{isotropia}}. \label{figuraIsotropica}
\end{figure}

The criteria for the physical admissibility are well known (as example see reference \cite{Lake}), and could represent a particular case of the conditions of subsection \ref{aceptabilidadanisotropia} when $p_r=p_t$ (excluding the energy momentum tensor conditions). 
In this work we will choose the following values of mass and radius: 
$\bar{M}=0,87$ solar mass and $R=7,866$ KM with a compactness factor $u=\bar{M}/R=0,16$; $\bar{M}=0,9041$ solar mass and $R=6$ KM with a compactness factor $u=\bar{M}/R=0,22$ and $\bar{M}=1,77$ solar mass and $R=9,56$ KM with a compactness factor $u=\bar{M}/R=0,27$ (these values are typical of compact objects. Indeed these correspond to the starts 4U 1538-52, RXJ 1856-37 and Vela X-1, respectively \cite{anisotropico51,anisotropico52}).
With these values we obtain the constants $a$, $c$ and $A$ from equations (\ref{a1}), (\ref{c1}) and (\ref{continuidadheintt}). In the figure (\ref{figuraIsotropica}) we observe that in the three cases the density and pressure are positives and decreasing, the pressure vanishes at the boundary and the square of light velocity is less than 1 (for simplicity in the graphic analysis we take $\kappa=1$). Therefore Heintzmann's solution is well behaved under our physical assumptions.
Throughout the text $u$ represents in all figures the ratio mass-radius.  

\section{Applying MGD to Heintzmann's solution} \label{MGD}

In this section we will apply the MGD method to Heintzmann's solution and we will obtain two new analytic solutions. Furthermore, we will test if these new solutions are well behaved from physical point of view, to do this we will use realistic values of mass and radii that correspond to the starts 4U 1538-52, RXJ 1856-37 and Vela X-1 \cite{anisotropico51,anisotropico52}.

From equation (\ref{heint}) we have:
\begin{eqnarray}\label{nu1}
e^{\nu(r)}&=&A^{2}\left(1+ar^{2}\right)^{3}\\\label{mu1}
\mu(r)&=&1-\frac{3ar^{2}}{2}\frac{1+c\left(1+4ar^{2}\right)^{-1/2}}{1+ar^{2}}.
\end{eqnarray}

Now we apply the MGD method to Heintzmann's solution (\ref{heint}). Following equations (\ref{mu})-(\ref{nu}) we get:  

\begin{eqnarray}\label{nu33}
e^{\nu(r)}&=&A^{2}\left(1+ar^{2}\right)^{3}\\\label{mu34}
e^{-\lambda(r)}&=&1-\frac{3ar^{2}}{2}\frac{1+c\left(1+ar^{2}\right)^{-1/2}}{1+ar^{2}}+\alpha f^{*}(r),
\end{eqnarray}
where $f^{*}(r)$ is determined imposing a mimic constraint on the density or pressure. Respect this, two choices leading to physically acceptable solutions are following mimic constraints \cite{Ovalle10,Ovalle2}:

\begin{equation}
 \Theta^{0}_{0}(r)=\bar{\rho}(r), \label{mimic1}  
\end{equation}
or
\begin{equation}
    \Theta^{1}_{1}(r)=\bar{p}(r). \label{mimic2}
\end{equation}

As soon as any mimic constraint (\ref{mimic1}) or (\ref{mimic2}) is imposed, $f^{*}(r)$ can be determined . Respect to the constant parameters $a$, $c$ and $A$, they are determined matching the expressions (\ref{nu33}) and (\ref{mu34}) with the corresponding Schwarszchild's exterior solution.

\begin{equation}\label{schwarzschild}
ds^{2}=-\left(1-\frac{2M}{r}\right)dt^{2}+\left(1-\frac{2M}{r}\right)^{-1}dr^{2}+r^{2}d\Omega^{2},    
\end{equation}

The first matching condition is given by equations (\ref{secondform}) and (\ref{pradialeff}), obtaining: 

\begin{equation}\label{saltopresion1}
 p_r(R)=0,
\end{equation}
so:
\begin{equation}
\bar{p}(R)-\alpha \Theta^1_1(R)=0. \label{matchingpressure1}
\end{equation}

The second matching condition is given by equation (\ref{firstform}), yielding to:

\begin{eqnarray}
A^{2}\left(1+aR^{2}\right)^{3}&=1-\frac{2M}{R}, \label{continuidadtt}
\end{eqnarray}
and,
\begin{eqnarray}
\left(1-\frac{3aR^{2}}{2}\frac{1+c\left(1+4aR^{2}\right)^{-1/2}}{1+aR^{2}}\right)+ \alpha f^*&=1-\frac{2M}{R}. \label{continuidadrr}
\end{eqnarray}
Equating equations (\ref{continuidadtt}) and (\ref{continuidadrr}) we obtain:

\begin{equation}
A^{2}\left(1+aR^{2}\right)^{3}=\left(1-\frac{3aR^{2}}{2}\frac{1+c\left(1+4aR^{2}\right)^{-1/2}}{1+aR^{2}} \right) + \alpha f^* \label{continuidadheint1}
\end{equation}

\subsection{Admissibility of the solution} \label{aceptabilidadanisotropia}
On the other hand when one considers an anisotropic matter distribution the basic requirements  change slightly, these are given by references \cite{Harko1,Harko2,Herrera1}:
\begin{enumerate}
    \item the density and pressure $p_r$ should be positive inside the star.
    \item the gradients $d\rho/dr$ , $dp_r/dr$ and $dp_t/dr$ should be negative.
    \item  inside the static configuration the speed of sound should be less than the speed of light, i.e. $0 \le dp_r/d\rho \le1$ and $0 \le dp_t/d\rho \le 1$.
    \item a physically reasonable energy-momentum tensor has to obey the conditions $\rho-p_r-2p_r \ge 0$ and $\rho+p_r+2p_t \ge 0$.
    \item the radial pressure must vanish but the tangential pressure may not vanish at the boundary $r = R$ of the sphere. However, the radial pressure is equal to the tangential pressure at the center of the fluid sphere.
\end{enumerate}

In this article we will analyze the above criteria.
However, some of these requirements are still in discussion,for example in \cite{Maartens} it has been claimed that a solution which disrupts the condition $3$ could still be allowed. 

Other authors argued that the speed of sound must decreases with increasing $r$. But the latest assumption could be not valid for anisotropic fluid, since the speed of sound at the isotropic case could be directly related to the material rigidity. In facts it is expected that stable configurations have a greater rigidity inside and that it decrease with increasing $r$, {\it i.e.} the sound speed should be decreasing as increasing $r$. But the latest assumption is not mandatory for anisotropic fluids \cite{Herrera2}. 

Respect to item 1, some authors have argued that if ordinary matter is present, described by a high density state equation, in which anisotropies naturally appear, one should impose the condition of the positivity of all pressures. However, for more exotic types of matter (dark  energy, or as in our case: scalar or vectorial fields, or generic gravitational sources, etc.) the tangential pressure may be negative inside the star, and it may not vanish on the surface. Exotic objects like compact gravastars or compact boson stars could have negative pressures \cite{Harko3} .

\subsection{Solution I: mimic constraint for pressure}
\label{constraint1}

We will use the contraint (\ref{mimic2}), so from equation (\ref{pradialeff}) we get the following value for the effective radial pressure:  

\begin{equation}\label{pradialeff1}
 p_{r}=(1-\alpha)\bar{p}(r),   
\end{equation}
where $\bar{p}(r)$ is given by equation (\ref{presionheint}).

Equating equations (\ref{pr}) and (\ref{dos}) we obtain the following expression for $f^{*}(r)$: 

\begin{eqnarray}\label{f*}
f^{*}(r)=\frac{3ar^{2}}{2}\frac{1+c\left(1+4ar^{2}\right)^{-1/2}}{1+ar^{2}}+\frac{ar^{2}+1}{7ar^{2}+1}-1.
\end{eqnarray}

Hence from equation (\ref{mu34}) the deformed radial metric component reads:
\begin{equation}\label{radialperturb}
e^{-\lambda(r)}=\left(1-\alpha\right)\mu(r)+\alpha\left(\frac{ar^{2}+1}{7ar^{2}+1}\right),
\end{equation}
where $\mu(r)$ corresponds to equation (\ref{mu1}).
The interior metric functions given by Eqs. (\ref{nu33}) and (\ref{radialperturb}) represent the Heintzmann's solution minimally deformed by the generic anisotropic source $\Theta_{\mu\nu}$. We can see that the limit $\alpha\rightarrow 0$ in Eq. (\ref{radialperturb}) leads to the standard Heintzmann's solution for perfect fluid.

At this point, we will match the deformed interior metric given by  (\ref{nu33}) and (\ref{radialperturb}),with the exterior vacuum Schwarzschild solution (\ref{schwarzschild}).
Using (\ref{saltopresion1}), and since $\alpha \neq 0$ in equation (\ref{pradialeff1}) we arrive to:
\begin{equation}
    \bar{p}(R)=0,
\end{equation}
which, according to the Eq. (\ref{presionheint}) leads to:
\begin{equation}\label{cmgd}
c=\frac{3\left(1+4aR^{2}\right)^{1/2}\left(1-aR^{2}\right)}{1+7aR^{2}}. 
\end{equation}

The continuity of the metric is given by the equations (\ref{continuidadtt}) and (\ref{continuidadrr}):
\begin{equation}\label{continuidadttmgd}
A^{2}\left(1+aR^{2}\right)^{3}=1-\frac{2M}{R},
\end{equation}
and
\begin{equation}\label{continuidadrrmgd}
\left(1-\alpha\right)\mu(r)+\alpha\left(\frac{ar^{2}+1}{7ar^{2}+1}\right)=1-\frac{2M}{R}, 
\end{equation}

Eqs.  (\ref{cmgd}), (\ref{continuidadttmgd}) and (\ref{continuidadrrmgd})  are the necessary and sufficient conditions to match the interior solution
with the exterior Schwarzschild space-time. 

We will evaluate $c$ from equation (\ref{cmgd}) at equation (\ref{continuidadrrmgd}) and we get the following expression for $a$: 
\begin{equation}
    a=\frac{M}{R^2(3R-7M)}. \label{a2}
\end{equation}
Then with above expression (\ref{a2}), we will obtain the value for $c$ from equation (\ref{cmgd}) and for $A$ from equation (\ref{continuidadttmgd}).

On the other hand replacing $f^{*}(r)$ given by (\ref{f*}) in the equations (\ref{cero}) and (\ref{tres}) we obtain $\Theta^{0}_{0}(r)$ and $\Theta^{2}_{2}(r)$:

\begin{eqnarray}
     \Theta^0_0(r)= -\frac{9a}{2\kappa^{2}} \frac{\zeta(r)}{\left(ar^{2}+1\right)^{2}\left(7ar^{2}+1\right)^{2}\left(4ar^{2}+1\right)^{3/2}} \\
     \Theta^2_2(r)= -\frac{3a}{2\kappa^{2}} \frac{\xi(r)}{\left(ar^{2}+1\right)^{2}\left(7ar^{2}+1\right)^{2}\left(4ar^{2}+1\right)^{1/2}},
\end{eqnarray}
where the functions $\zeta(r)$ and $\xi(r)$ are:

\begin{eqnarray}
    \zeta(r)=& \sqrt{1+4ar^2} \Big (-3-15ar^2+19a^2r^4+131a^3r^6+28a^4r^8 \Big)+17car^2+91a^2r^4c \nonumber \\
    &+147a^3r^6c+c \\
    \xi(r)=&\sqrt{1+4ar^2} \Big (189 a^3r^6-45a^2r^4-21ar^2-3\Big)+343a^3r^6c+147a^2r^4c+21car^2+c.
\end{eqnarray}

The above expressions allow to us find the effective density and the effective tangential pressure from equations (\ref{rhoeff}) and (\ref{ptangencialeff}):
\begin{eqnarray}\label{denperb}
{\rho}(r)&=&\bar{\rho}(r)-\frac{\alpha}{2\kappa^{2}} \frac{9a\zeta(r)}{\left(ar^{2}+1\right)^{2}\left(7ar^{2}+1\right)^{2}\left(4ar^{2}+1\right)^{3/2}}\\\label{tanperb}
{p}_{t}(r)&=&\bar{p}(r)+\frac{\alpha}{2\kappa^{2}} \frac{3a\xi(r)}{\left(ar^{2}+1\right)^{2}\left(7ar^{2}+1\right)^{2}\left(4ar^{2}+1\right)^{1/2}}.
\end{eqnarray}

\begin{figure}[H]
\centering 
\includegraphics [width=7.0 in]{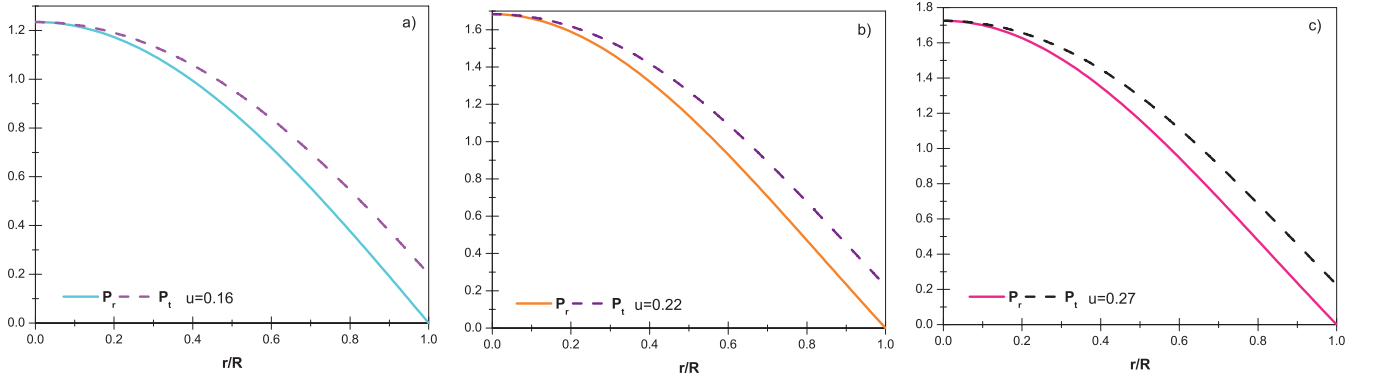}
\caption{The three panels display the radial (lower curve) and the tangential (upper curve) pressures, exhibiting a monotonically decreasing behaviour with the increasing $r$. All the above quantities are divided by $a$, {\it i.e.} $P_r=p_r/a$ and $P_t=p_t/a$.These plots correspond to solution of subsection \ref{constraint1} with $\alpha=0.2$.} \label{figuraPresionesc1}
\end{figure}

\begin{figure}[H]
\centering 
\includegraphics [width=7.0 in]{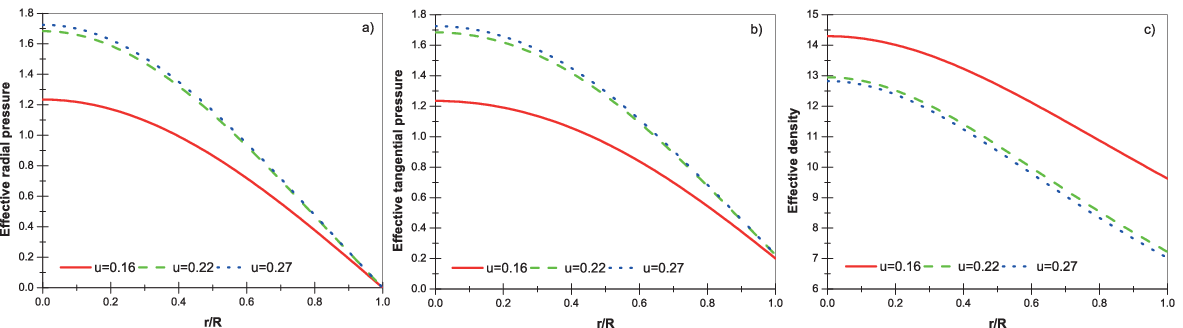}
\caption{Panel $a)$ shows the effective radial pressure. Panel $b)$ shows the effective tangential pressure and panel $c)$ shows the decreasing effective energy density. All these quantities are normalized by $a$. These plots correspond to solution of subsection \ref{constraint1} with $\alpha=0.2$.} \label{figuraPresionesyDensityc1}
\end{figure}

\begin{figure}[H]
\centering 
\includegraphics [width=5.5 in]{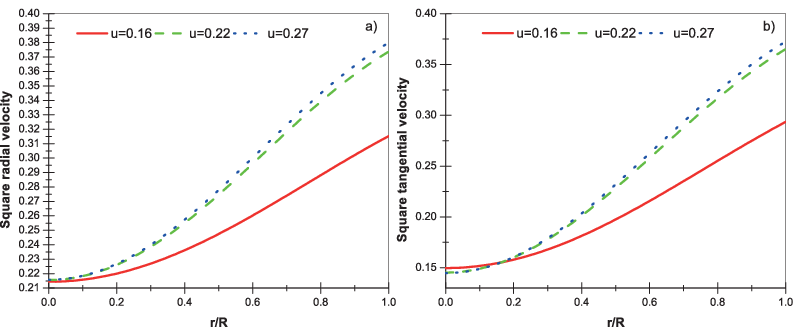}
\caption{Panels $a)$ and $b)$ show the squares of radial and tangential velocities, respectively. These plots correspond to solution of subsection \ref{constraint1} with $\alpha=0.2$.} \label{figuraVelocidadc1}
\end{figure}

\begin{figure}[H]
\centering 
\includegraphics [width=5.5 in]{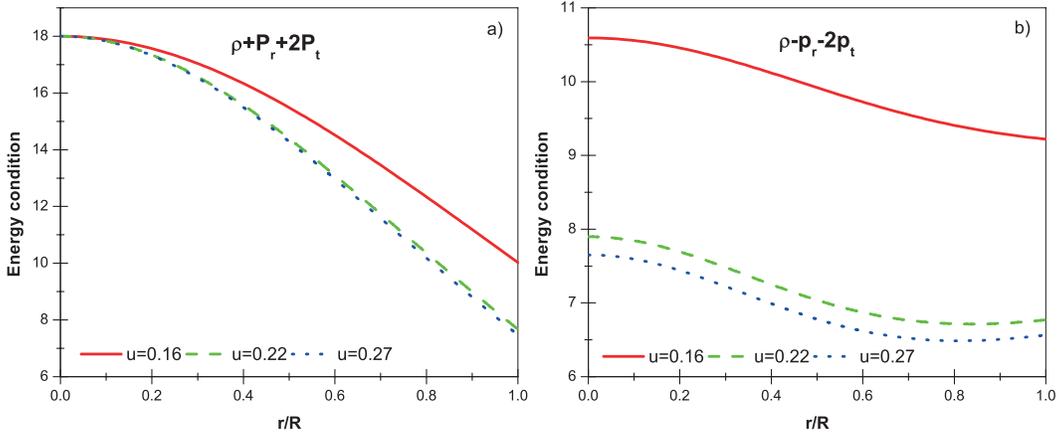}
\caption{Panel $a)$ and $b)$ show the energy conditions $\rho+P_r+2P_t$ and $\rho-P_r-2P_t$ respectively. These plots correspond to solution of subsection \ref{constraint1} with $\alpha=0.2$.} \label{figuraEMc1}
\end{figure}

From the physical admissibility analysis we will fix as free parameters the constants $M$ and $R$ corresponding again to the same values of mass and radius above declared in section \ref{isotropia}. We evaluate at equation (\ref{a2}) and obtain the value of $a$ and from equation (\ref{cmgd}) we obtain the value of $c$. In figures (\ref{figuraPresionesc1}), (\ref{figuraPresionesyDensityc1}), (\ref{figuraVelocidadc1}) and (\ref{figuraEMc1}) we see that all conditions from physical admissibility of subsection (\ref{aceptabilidadanisotropia}) are satisfied.

\subsection{Solution II: mimic constraint for density}
\label{constraint2}

An alternative choice leading to a physically acceptable solution is the “mimic constraint” for the density (\ref{mimic1}), then from equation (\ref{rhoeff}) we obtain:

\begin{equation}
    \rho= \bar{\rho} (1+\alpha),
\end{equation}
then equating the equations (\ref{ro}) and (\ref{cero}) we get $f^*(r)$, which reads:

\begin{equation}
f^{*}(r)=\frac{B}{r}-\frac{3ar^{2}}{2}\frac{1+c\left(1+4ar^{2}\right)^{-1/2}}{1+ar^{2}}, \label{f2}
\end{equation}
where the density $\bar{\rho}$ from Eq. (\ref{densidadheint}) has been used. To avoid a singular behaviour at the center $r=0$ we must impose $B=0$, then the equation (\ref{mu}) yields:
\begin{equation}\label{lambmimicden}
e^{-\lambda(r)}=\mu(r)-\alpha\left(\frac{3ar^{2}}{2}\frac{1+c\left(1+4ar^{2}\right)^{-1/2}}{1+ar^{2}}\right).
\end{equation}

Next evaluating $f^*(r)$ from equation (\ref{f2}) at equations (\ref{dos}) and (\ref{tres}) we find the expressions for $\Theta^1_1(r)$ and $\Theta^2_2(r)$:

\begin{eqnarray}
    \Theta^1_1=\frac{3a}{2\kappa^2} \frac{(\sqrt{1+4ar^2}+c)(1+7ar^2)}{(1+ar^2)^2 \sqrt{1+4ar^2}} \label{theta1contraint2} \\
    \Theta^2_2=\frac{3a}{2\kappa^2} \frac{\sqrt{1+4ar^2}(9ar^2+1)+7ar^2c+c}{(1+ar^2)^2 \sqrt{1+4ar^2}}. \label{theta2contraint2}
\end{eqnarray}

With these expressions  (\ref{theta1contraint2}) and (\ref{theta2contraint2}) the radial and tangential pressures (\ref{pradialeff}) and (\ref{ptangencialeff}) are:
\begin{equation}
{p}_{r}(r)=\bar{p}(r)-\alpha\frac{3a}{2\kappa^2} \frac{(\sqrt{1+4ar^2}+c)(1+7ar^2)}{(1+ar^2)^2 \sqrt{1+4ar^2}}, \label{pr2}
\end{equation}
and
\begin{equation}
{p}_{t}(r)=\bar{p}(r)-\alpha\frac{3a}{2\kappa^2} \frac{\left(\sqrt{1+4ar^2}(9ar^2+1)+\left(7ar^2+1\right)c\right)}{(1+ar^2)^2 \sqrt{1+4ar^2}}.
\end{equation}

The continuity of the effective radial pressure (\ref{saltopresion1}) along with the equation (\ref{pr2}), yield to:
\begin{equation}\label{cmimicden}
c=-\frac{\sqrt{1+4aR^2}(-3+3aR^2+\alpha+7\alpha aR^2)}{(1+\alpha)(1+7aR^2)}.
\end{equation}.

Using the expression given by (\ref{f2}), the matching conditions in (\ref{continuidadtt}) and (\ref{continuidadrr}) lead to:
\begin{equation}\label{Amimicden}
A^{2}\left(1+aR^{2}\right)^{3}=1-\frac{2M}{R},    
\end{equation}
and:
\begin{equation}\label{amimicden}
1-\left(1+\alpha\right)\left(\frac{3ar^{2}}{2}\frac{1+c\left(1+4ar^{2}\right)^{-1/2}}{1+ar^{2}}\right)=1-\frac{2M}{R},
\end{equation}

Eqs. (\ref{cmimicden})-(\ref{amimicden}) are the necessary and sufficient conditions to match the exterior Schwarzschild
solution with the deformed interior solution.

Inserting $f^*(r)$ from equation (\ref{f2}) and $c$ from equation (\ref{cmimicden}) at equation (\ref{amimicden}) we again find the same value for $a$ from equation (\ref{a2}). Next with this value for $a$ we will obtain the value for $c$ from equation (\ref{cmimicden}) and A from equation (\ref{Amimicden}).\\

\begin{figure}
\centering 
\includegraphics [width=7.0 in]{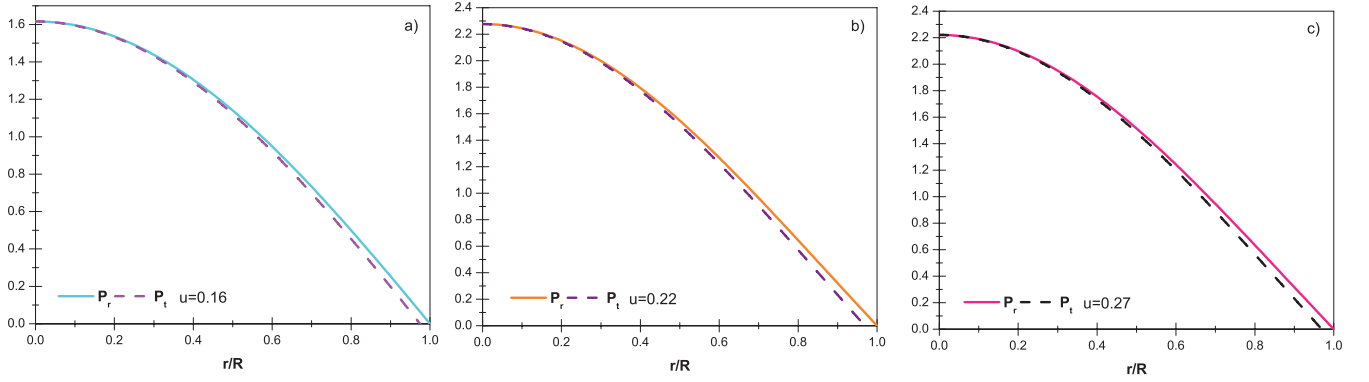}
\caption{The three panels display the tangential (lower curve) and the radial (upper curve) pressures, exhibiting a monotonically decreasing behaviour with the increasing $r$. All the above quantities are divided by the constant $a$, {\it i.e.} $P_r=p_r/a$ and $P_t=p_t/a$. These plots correspond to solution of subsection \ref{constraint2} with $\alpha=0.3$.} \label{figuraPresionesc2alpha+}
\end{figure}

\begin{figure}[H]
\centering 
\includegraphics [width=7.0 in]{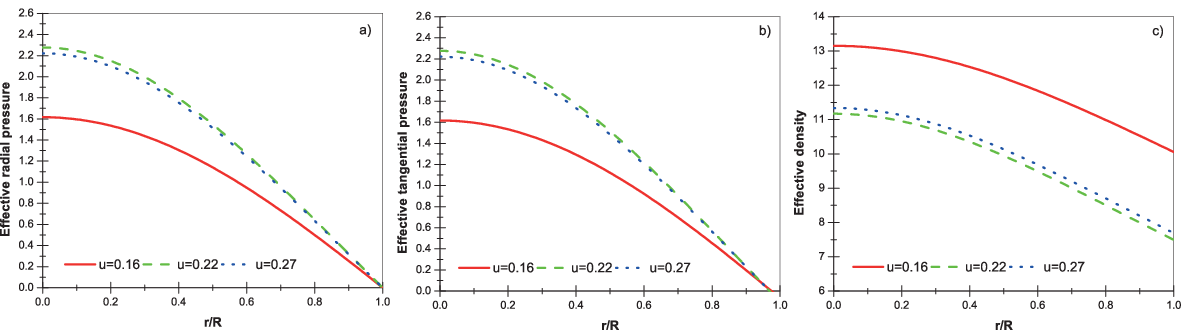}
\caption{Panel $a)$ shows the effective radial pressure. Panel $b)$ shows the effective tangential pressure and panel $c)$ shows the decreasing effective energy density. All these quantities are divided by the value of $a$. These plots correspond to solution of subsection \ref{constraint2} with $\alpha=0.3$.}\label{figuraPresiondensityc2alpha+}
\end{figure}

\begin{figure}[H]
\centering 
\includegraphics [width=5.5 in]{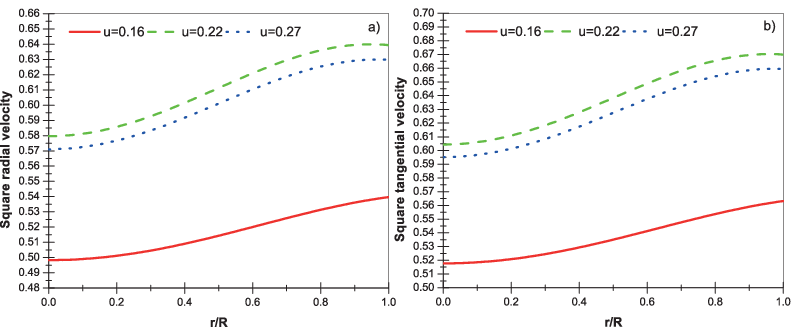}
\caption{Panels $a)$ and $b)$ show the squares of radial and tangential velocities respectively. These plots correspond to solution of subsection \ref{constraint2} with $\alpha=0.3$.} \label{figuraVelc2alpha+}
\end{figure}

\begin{figure}
\centering 
\includegraphics [width=6.5 in]{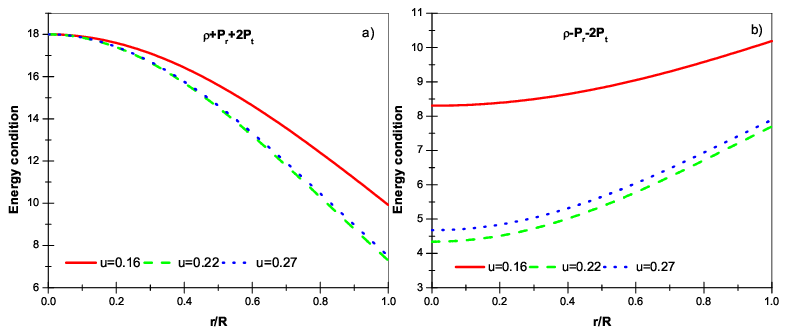}
\caption{Panel $a)$ and $b)$ show the energy conditions $\rho+P_r+2P_t$ and $\rho-P_r-2P_t$, respectively. These plots correspond to solution of subsection \ref{constraint2} with $\alpha=0.3$.} \label{figuraEMc2alpha+}
\end{figure}

For testing the admissibility of our solution we again choose the same values of mass and radius from subsections \ref{isotropia} and \ref{constraint1} but now $\alpha=0,3$, obtaining $a$ from equation (\ref{a2}) and $c$ from equation (\ref{cmimicden}).
For these values figure ($\ref{figuraPresionesc2alpha+}$) shows the radial pressure ${p}_{r}(r)$ and tangential pressure ${p}_{t}(r)$ inside the spherical distribution, showing how the magnitude of the anisotropy increases towards the surface.The tangential pressure is less than radial pressure {\it i.e} there is an attractive force. Also we observe that the tangential pressure becomes negative before the boundary. In figures (\ref{figuraPresionesc2alpha+}),(\ref{figuraPresiondensityc2alpha+}),(\ref{figuraVelc2alpha+}) and (\ref{figuraEMc2alpha+}) we observe that all requirements of admissibility of subsection (\ref{aceptabilidadanisotropia}) are fulfills . 

\begin{figure}
\centering 
\includegraphics [width=7.0 in]{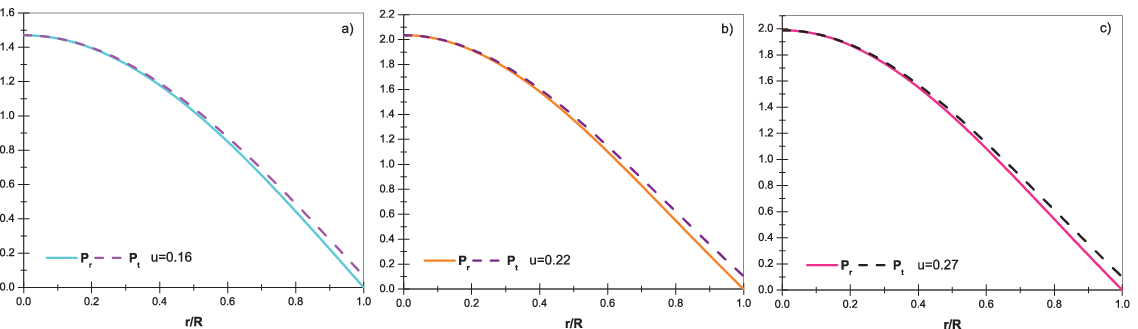}
\caption{The three panels display the radial (lower curve) and the tangential (upper curve) pressures, exhibiting a monotonically decreasing behaviour with the increasing $r$. All the above quantities are divided by the constant $a$, {\it i.e.} $P_r=p_r/a$ and $P_t=p_t/a$. These plots correspond to solution of subsection \ref{constraint2} with $\alpha=-0.3$.} \label{figuraPresionesc2alpha-}
\end{figure}

\begin{figure}
\centering 
\includegraphics [width=7.0 in]{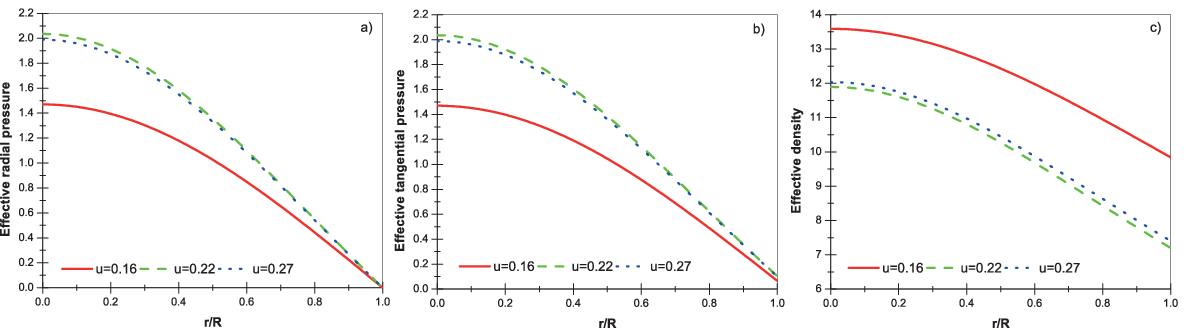}
\caption{Panel $a)$ shows the effective radial pressure. Panel $b)$ shows the effective tangential pressure and panel $c)$ shows the decreasing effective energy density. All these quantities are divided by the value of $a$. These plots correspond to solution of subsection \ref{constraint2} with $\alpha=-0.3$.}\label{figuraPresiondensityc2alpha-}
\end{figure}

\begin{figure}
\centering 
\includegraphics [width=5.5 in]{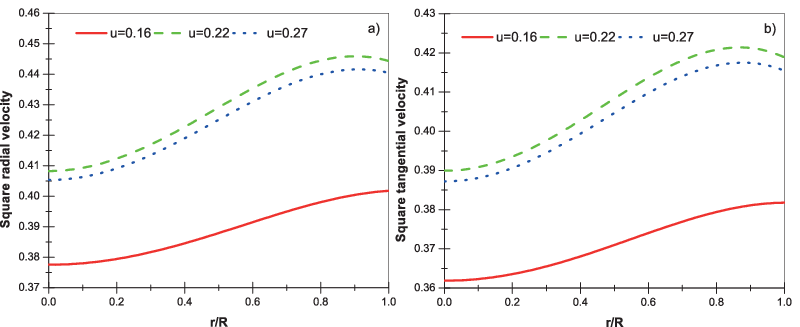}
\caption{Panels $a)$ and $b)$ show the squares of radial and tangential velocities, respectively. These plots correspond to solution of subsection \ref{constraint2} with $\alpha=-0.3$.} \label{figuraVelc2alpha-}
\end{figure}

\begin{figure}
\centering 
\includegraphics [width=5.5 in]{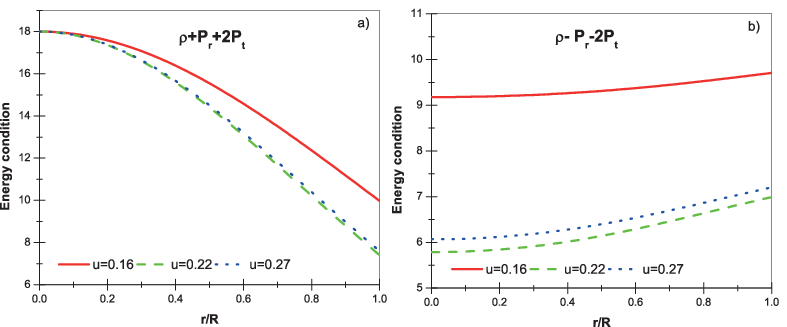}
\caption{Panel $a)$ and $b)$ show the energy conditions $\rho+P_r+2P_t$ and $\rho-P_r-2P_t$, respectively. These plots correspond to solution of subsection \ref{constraint2} with $\alpha=-0.3$.} \label{figuraEMc2alpha-}
\end{figure}

With the same values of mass and radius also it is possible show a scenario where the tangential pressure is greater than radial pressure and being both positives {\i.e} there is a repulsive force, for example with $\alpha=-0,3$. In figures (\ref{figuraPresionesc2alpha-}), (\ref{figuraPresiondensityc2alpha-}), (\ref{figuraVelc2alpha-}) and (\ref{figuraEMc2alpha-}) we see  again, that all the admissibility conditions are fulfills, therefore with  $\alpha=-0,3$ the solution is also well behaved.

\subsection{Stability conditions}

\begin{figure}
\centering 
\subfigure[Solutions of subsection \ref{constraint1}.]{\includegraphics[width=80mm]{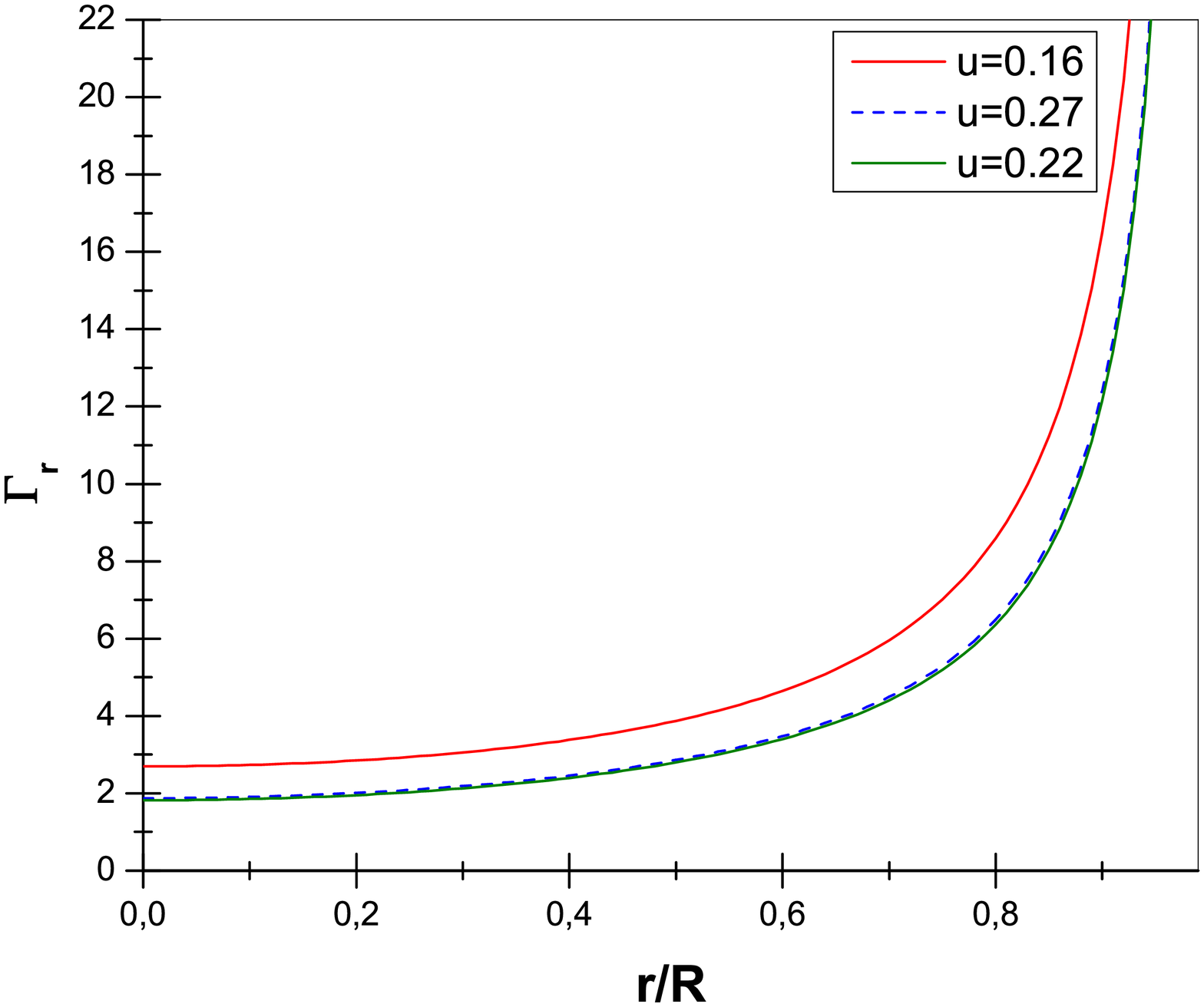}} \label{B1}
\subfigure[Solutions of subsection \ref{constraint2} with $\alpha=-0.3$]{\includegraphics[width=80mm]{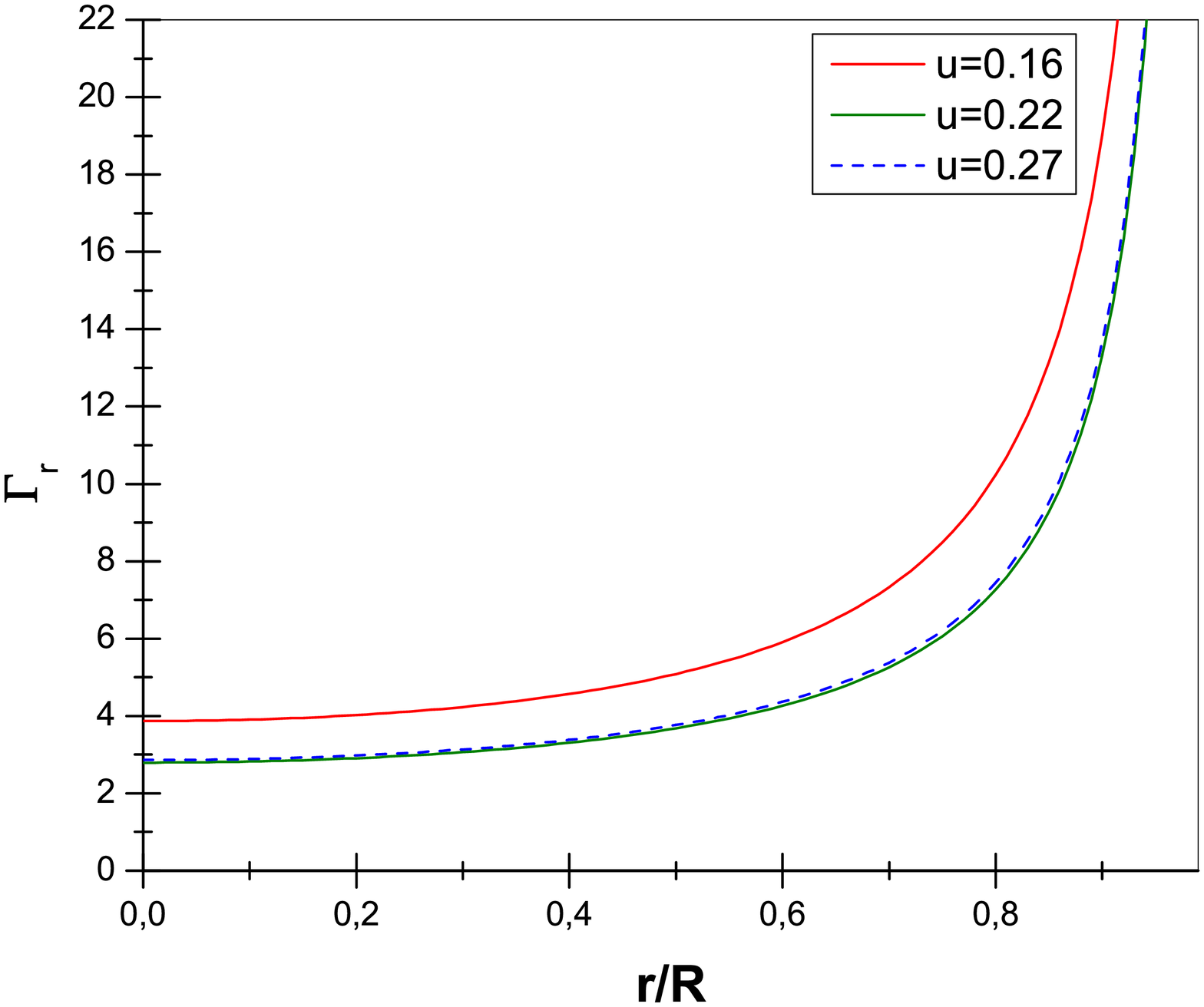}} \label{B2}
\subfigure[Solutions of subsection \ref{constraint2} with $\alpha=0.3$]{\includegraphics[width=80mm]{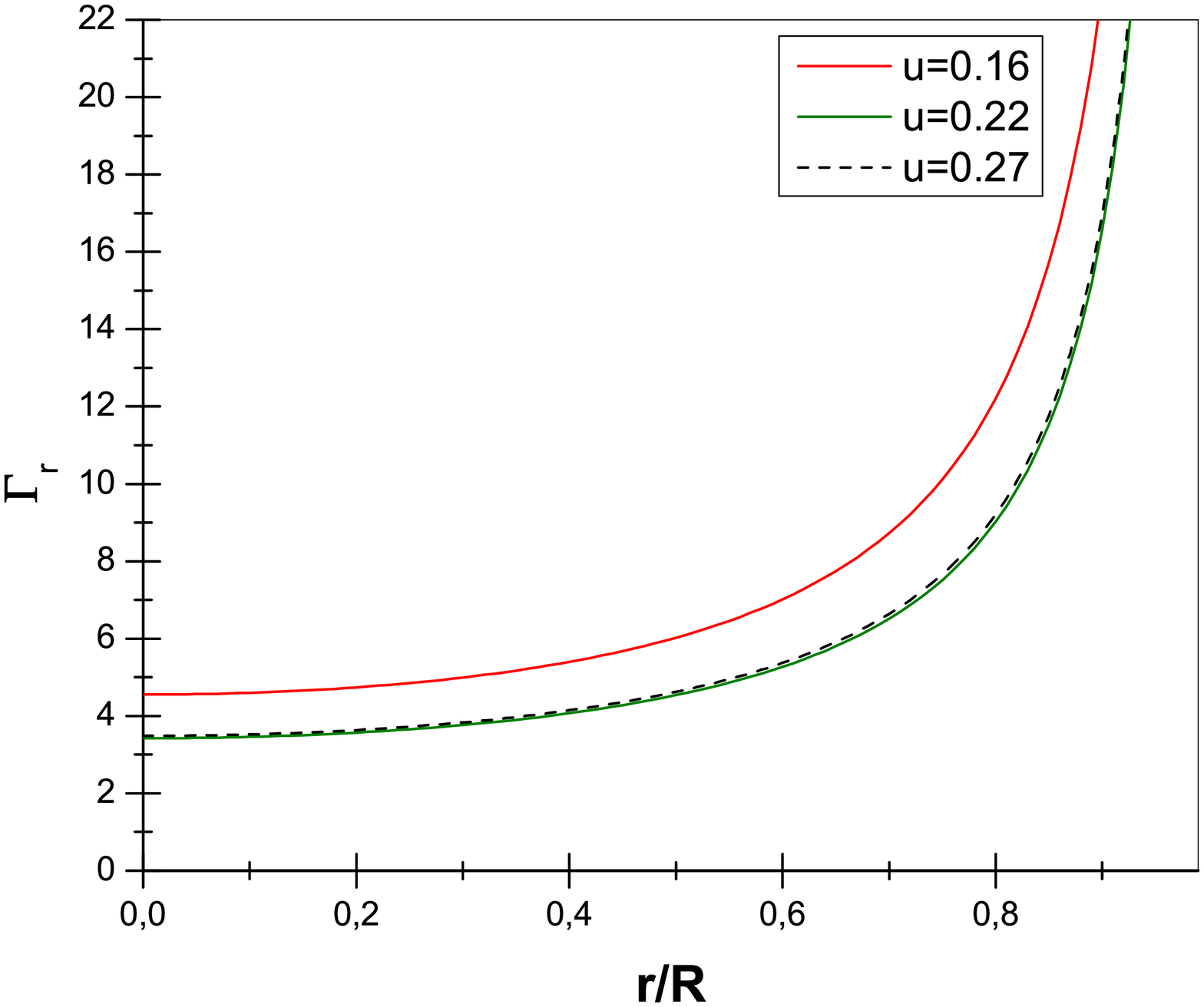}} \label{B3}
\caption{Adiabatic index criterion.} 
\label{IndiceAdiabaticoCriterio}
\end{figure}

\begin{figure}
\centering 
\subfigure[Solutions of subsection \ref{constraint1}.]{\includegraphics[width=80mm]{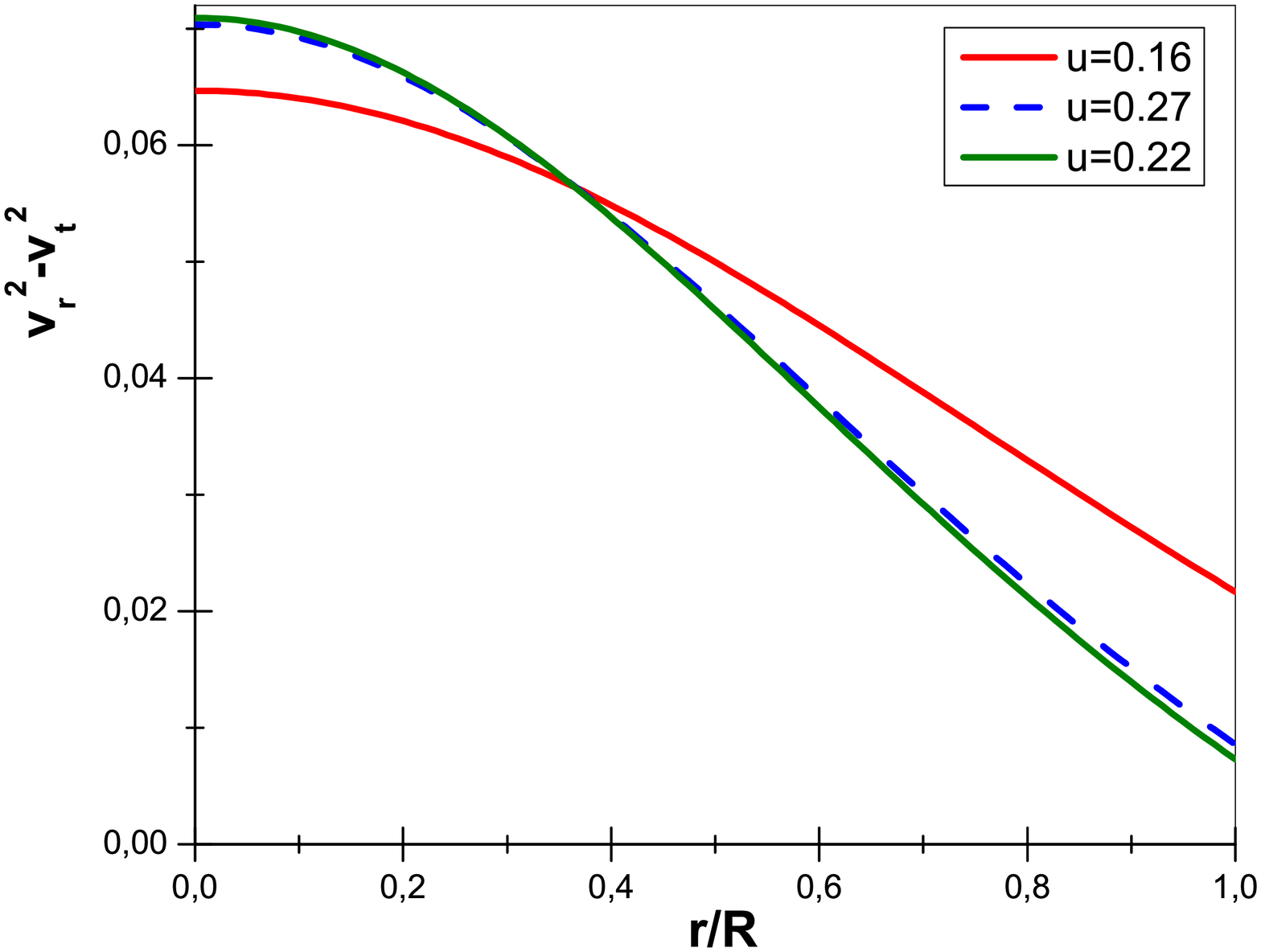}} \label{A1}
\subfigure[Solutions of subsection \ref{constraint2} with $\alpha=-0.3$]{\includegraphics[width=80mm]{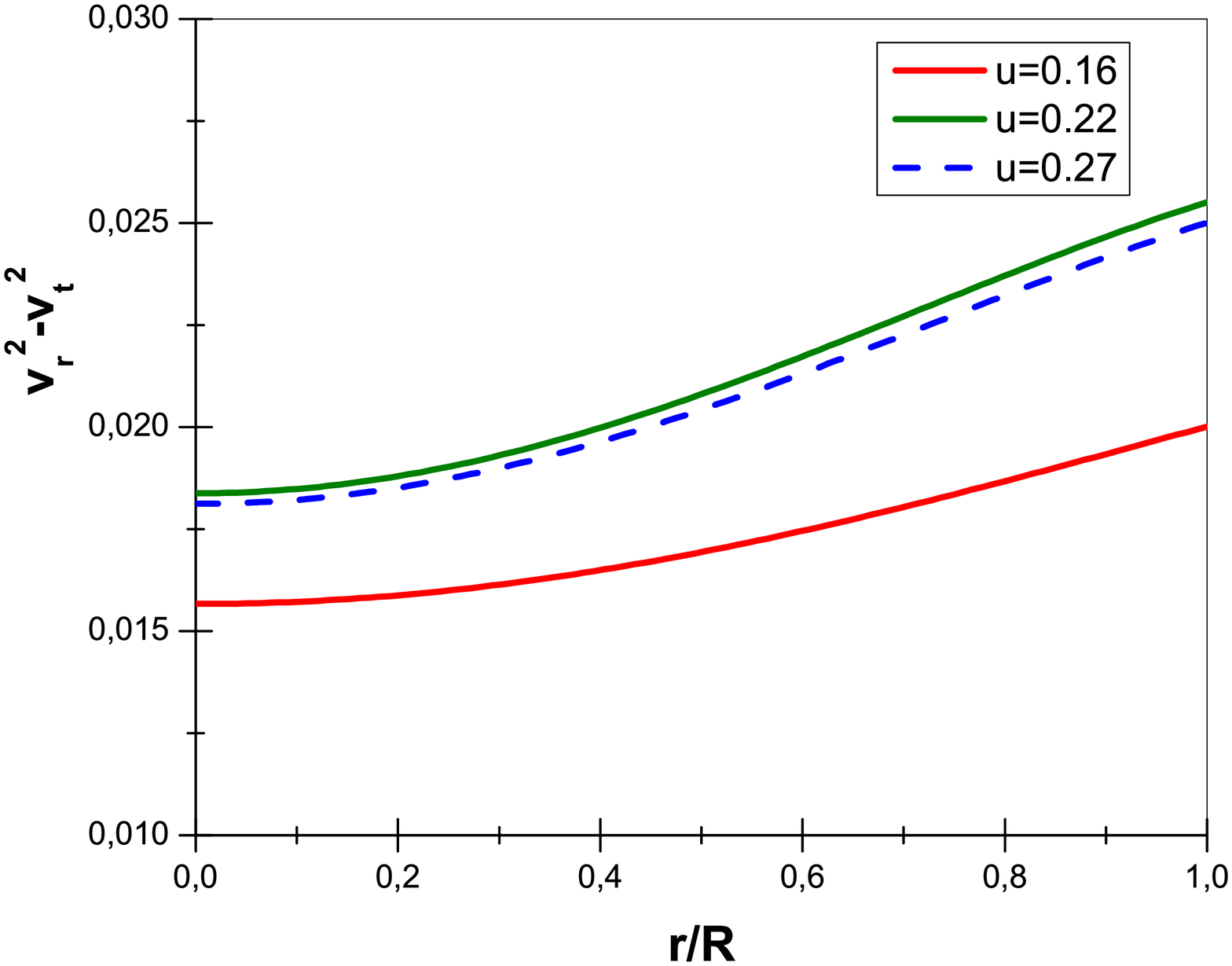}} \label{A2}
\subfigure[Solutions of subsection \ref{constraint2} with $\alpha=0.3$]{\includegraphics[width=80mm]{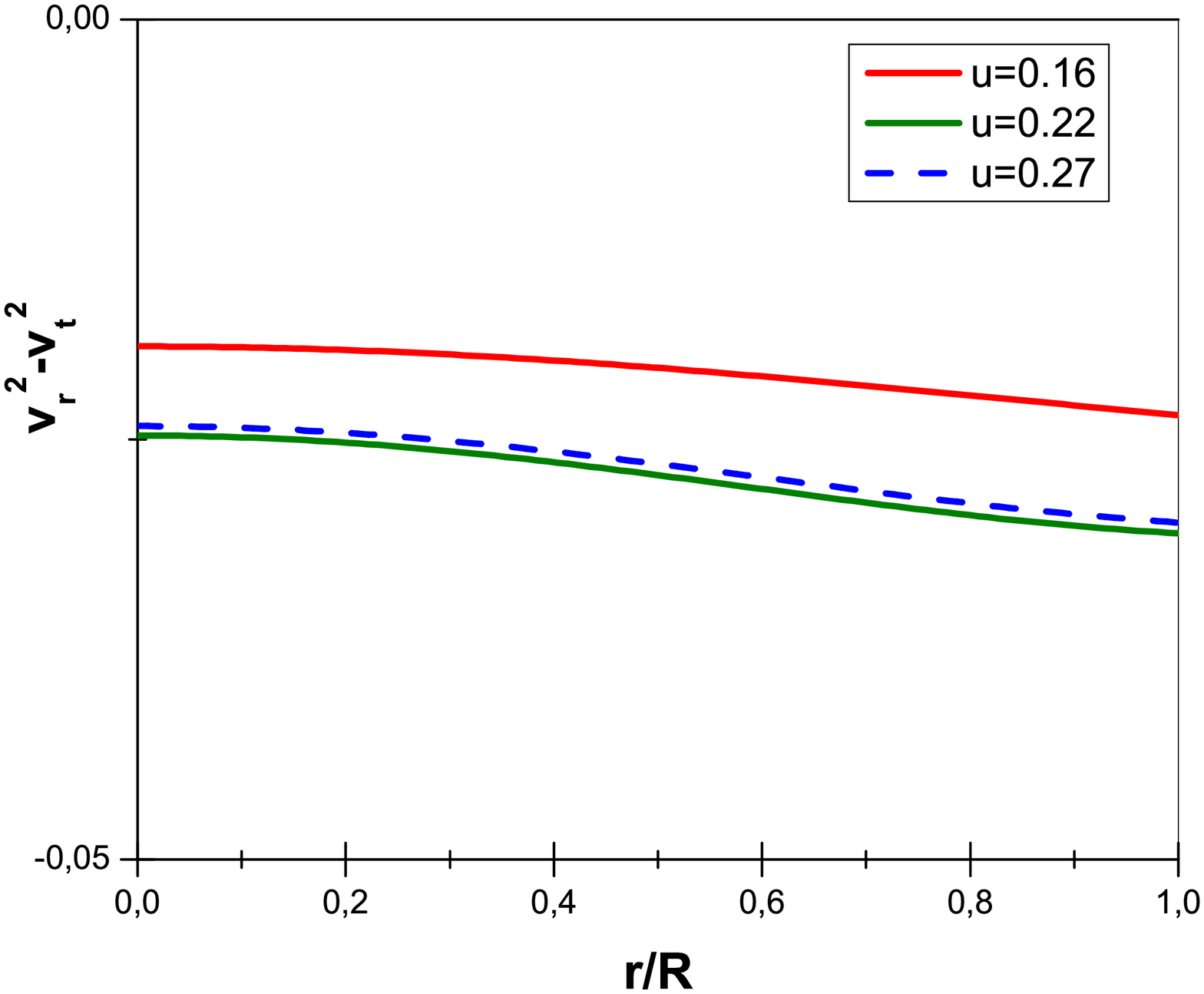}} \label{A3}
\caption{Abreu criterion $v_r^2-v_t^2$.} 
\label{AbreuCriterio}
\end{figure}

One important aspect in the study of compact objects in general relativity describing anisotropic matter distributions is the stability of the model. In this section we analyze the stability of the present model from two perspectives. The first one is via the adiabatic index $\Gamma$ and the second one is through the analysis of the square of the sound speeds using Abreu et. al.\cite{Abreu} studies based on the Herrera's cracking concept \cite{herrera5}. \\
It is well known that a spherically symmetric and static configuration associated with a Newtonian isotropic matter distribution will collapse if the adiabatic index is $\Gamma<4/3$. In distinction with the relativistic anisotropic fluid spheres the collapsing condition becomes \cite{Herrera3,Herrera4} 
\begin{equation}\label{adibatic}
\Gamma<\frac{4}{3}+\left[\frac{1}{3}\kappa\frac{\rho_{0}p_{r0}}{|p^{\prime}_{r0}|}r+\frac{4}{3}\frac{\left(p_{t0}-p_{r0}\right)}{|p^{\prime}_{r0}|r}\right]_{max}    
\end{equation}
where $\rho_{0}$, $p_{r0}$ and $p_{t0}$ are the initial density, radial and tangential pressure when the fluid is in static equilibrium. The second term in the right hand side represents the relativistic corrections to the Newtonian perfect fluid and the third term is the contribution due to anisotropy. Heintzmann and Hillebrandt \cite{hillebrandt} showed that the stability condition for a relativistic compact object is given by $\Gamma>4/3$. Since the gravitational collapse takes place in the radial direction of the configuration it is sufficient to calculate the adiabatic index $\Gamma_{r}$ in such direction. We can explicitly obtain the adiabatic index from,  \cite{chan}
\begin{equation}
\Gamma_{r}=\frac{\rho+p_{r}}{p_{r}}\frac{dp_{r}}{d\rho}.    
\end{equation}
Figure (\ref{IndiceAdiabaticoCriterio}) shows that all our models satisfies the above criteria. Therefore it is a stable model. On the other hand, Abreu et. al. analysis regarding the square of the sound speeds within the stellar configuration, provides us a criteria to determine the regions where the model is stable. So the model is stable if $v^{2}_{r}-v^{2}_{t}$ remains between $0$ and $1$ and unstable if  $v^{2}_{r}-v^{2}_{t}$ remains between $-1$ and $0$.
We can see that solutions of figures \ref{AbreuCriterio}(a) and \ref{AbreuCriterio}(b) becomes potentially stable since they fulfill the condition $0<v_r^2-v_t^2<1$, whereas the solutions of figure \ref{AbreuCriterio}(c) becomes unstable.

\section{Conclusions and remarks}
We showed that by imposing our matching conditions the isotropic Heintzmann's solution is well behaved using typical mass and radius values for some compact objects such as 4U 1538-52, RXJ 1856-37 and Vela X-1,  whose compactness factor $u$ are $0,16$, $0,22$ and $0,27$, respectively. The choice of Heintzmann IIa solution is merely arbitrary, because there is not evidence that the mentioned compact stars are described fully by this solution.

Then, applying the Minimal Geometric Deformation method to Heintzmann's solution we have found two new analytic anisotropic and spherically symmetric solutions of Einstein field equations. These two solutions are the results of the decoupling of Einstein equations in a isotropic sector described by $\bar{T}_{\mu \nu}$ and the sector described by the quasi-Einstein equations with a source $\Theta_{\mu \nu}$. Due to the application of MGD the perfect and the extra fluid are separately conserved. So, the combination of these two sectors has only gravitational interaction, and does not has exchange of energy momentum.

We have shown that these two new solutions are well behaved for the above parameters of mass and radius, since they fulfill every admissibility criteria of the subsection (\ref{aceptabilidadanisotropia}). To accomplish it we use $\alpha=0,2$ and $\alpha=0,3$ in the first and second constraint, respectively. Moreover considering $\alpha=0,3$ in the second constraint, we obtained an attractive force where the tangential pressure becomes negative before the boundary of the compact object, but using $\alpha=-0,3$ a repulsive force appears. In all cases the solutions are well behaved. The matching conditions are perfectly satisfied in both solutions.

All the sources $\Theta_{\mu \nu}$ used in this works are generic gravitational sources, therefore does not represent to a high density equation of state {\it i.e.} does not impose that the tangential pressure is positive inside of the fluid sphere. 

\begin{figure}
\centering 
\includegraphics [width=4.0 in]{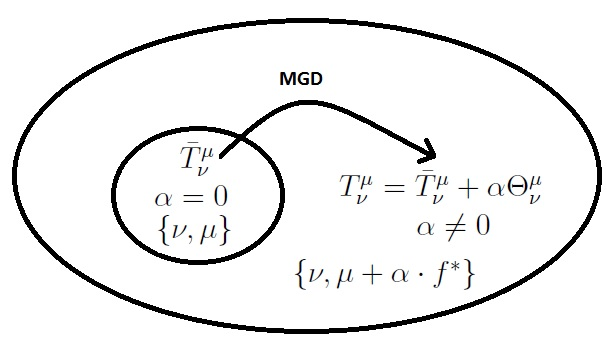}
\caption{Minimal Geometric Deformation approach. } \label{esquema}
\end{figure}

In the figure (\ref{esquema}) we observe how any perfect fluid solution can be consistently extended to the anisotropic domain via the MGD approach. In our case the seed well behaved Heintzmann's solution $\{\nu,\mu \}$ is extended to an anisotropic well behaved scenario $\{ \nu, \mu +\alpha \cdot f^* \}$.

In this work we do not make a complete study about compact objects, however our solutions would serve for future studies about these objects. Moreover, working in CGS system table \ref{tabla1} shows the values of the central and surface energy density and central pressure, corresponding to the isotropic Heintzmann's solution from section \ref{isotropia} and both anisotropic extensions from  subsection \ref{constraint1} (mimic constraint for pressure) and from subsection \ref{constraint2} (mimic constraint for density). We found values for central and surface density of order of $10^{14}-10^{15}g/cm^3$ and central pressure of order of $10^{34}-10^{35}$ $\mbox{dyne}/cm^2$. These are typical values for compact objects, some examples can be found in references  \cite{anisotropico51,anisotropico52,Anisotropico9}.

\begin{table}
\centering
\caption{Energy density and central pressure in CGS system.}
{\begin{tabular}{@{}lccccccr@{}} \noalign{\smallskip}\hline\noalign{\smallskip}
Solution & $u$ & $\alpha$ & $a$  & $c$ (dimension & Central den-& Surface den-& Central pressu-
\\
&  & &$(cm^{-2})$& less &sity ($g/cm^3$)&sity ($g/cm^3$)& re $(\mbox{dyne}/cm^2)$ \\ \noalign{\smallskip}\hline\noalign{\smallskip}

section \ref{isotropia} & $0,163$ & $0$ & $1,422 \cdot 10^{-13}$ & $1,969$ & $1,018\cdot 10^{15}$ &$7,57 \cdot 10^{14}$ & $1,06 \cdot 10^{35}$ \\
  & $0,222$ & $0$ & $4,287 \cdot 10^{-13}$ & $1,551$ & $2,634 \cdot 10^{15}$ &$1,672 \cdot 10^{15}$ & $4,49 \cdot 10^{35}$\\ 
  & $0,27$& $0$ & $2,756 \cdot 10^{-13}$ & $1,151$ & $1,43 \cdot 10^{15}$ &$7,612 \cdot 10^{14}$ & $3,683 \cdot 10^{35}$  \\ \noalign{\smallskip}\hline\noalign{\smallskip}

subsection \ref{constraint1} & $0,163$ &$0,2$ &  $1,422 \cdot 10^{-13}$    & $1,969$    & $1,089 \cdot 10^{15}$ & $7,324 \cdot 10^{14}$ &  $8,48 \cdot 10^{34}$\\
  &  $0,222$ & $0,2$ & $4,287 \cdot 10^{-13}$ & $1,551$ & $2,934 \cdot 10^{15}$ &$1,598 \cdot 10^{15}$ & $3,592 \cdot 10^{35}$\\ 
  & $0,27$& $0,2$ & $2,756 \cdot 10^{-13}$ & $1,151$ & $1,676 \cdot 10^{15}$ &$7,195 \cdot 10^{14}$ & $2,946 \cdot 10^{35}$ \\ \noalign{\smallskip}\hline\noalign{\smallskip}

subsection \ref{constraint2} & $0,163$ &$0,3$  & $1,422 \cdot 10^{-13}$     & $1,246$ & $1,001 \cdot 10^{15}$ & $7,653 \cdot 10^{14}$ & $5,551 \cdot 10^{34}$  \\   & $0,222$ & $0,3$ & $4,287 \cdot 10^{-13}$ & $0,9$ & $2,552 \cdot 10^{15}$ &$1,706 \cdot 10^{15}$ & $2,371 \cdot 10^{35}$\\ 
  & $0,27$& $0,3$ & $2,756 \cdot 10^{-13}$ & $0,558$ & $1,348 \cdot 10^{15}$ &$7,878 \cdot 10^{14}$ & $1,966 \cdot 10^{35}$\\ 
  & $0,163$  &$-0,3$  & $1,422 \cdot 10^{-13}$     & $3,311$ & $1.035 \cdot 10^{15}$ & $7,489 \cdot 10^{14}$ & $5,049 \cdot 10^{34}$  \\
  & $0,222$  &$-0,3$  & $4,287 \cdot 10^{-13}$     & $2,761$ & $2,722 \cdot 10^{15}$ & $1,638 \cdot 10^{15}$ & $2,119 \cdot 10^{35}$  \\
  & $0,27$  &$-0,3$  & $2,756 \cdot 10^{-13}$     & $2,251$ & $1,513 \cdot 10^{15}$ & $7,345 \cdot 10^{14}$ & $1,717 \cdot 10^{35}$  \\ \noalign{\smallskip}\hline\noalign{\smallskip}
\end{tabular} \label{tabla1}}
\end{table}

%%%%%%%%%%%%%%%%%%%%%%%%%%%%%%%%%%%%%%%%%%%%%%%%%%%%%%%%%%%%%%%%%%%%%%

\begin{acknowledgements}
 The work of ME  was partially founded by FONDECYT Regular under Grant No. 1151107 . FTO thanks the financial support of the post-graduate program at the Universidad de Antofagasta-Chile.
\end{acknowledgements}

%%%%%%%%%%%%%%%%%%%%%%%%%%%%%%%%%%%%%%%%%%%%%%%%%%%%%%%%%%%%%%%%%%%%%%%


\begin{thebibliography}{5}
\bibitem{Harko2}  M. K. Mak and T. Harko, {\it Chin. J. Astron. Astrophys.} {\bf 2} (2002) 248.

\bibitem{Harko1}  M. K. Mak, T. Harko , Anisotropic Stars in General Relativity, in {\it Proc.Roy.Soc.Lond.}, (2003) ,p.~393-408

\bibitem{anisotropico1} Singh, K.N., Pant, N. \& Govender, M ,
{\it Indian J Phys} , {\bf 90}, (2016) 1215.

\bibitem{anisotropico2} S.K. Maurya, Y.K. Gupta, Saibal Ray, Debabrata Deb , {\it Eur. Phys. J. C  } {\bf 76} (2016) 693.

\bibitem{anisotropico3} Maurya, S.K , {\it Eur. Phys. J. A} {\bf 53} (2017) 89.

\bibitem{anisotropico4} B.V.Ivanov , {\it Eur. Phys. J. C} {\bf 77 } (2017) 738

\bibitem{anisotropico5} Jasim, M.K., Maurya, S.K., Gupta, Y.K.  , {\it Astrophys Space Sci} {\bf 361 } (2016) 352.

\bibitem{anisotropico51} S.K. Maurya, Y.K. Gupta, Baiju Dayanandan, Saibal Ray, {\it Eur. Phys. J. C} {\bf 76} (2016)  266.  

\bibitem{anisotropico52} Piyali Bhar, Ksh. Newton Singh, Tuhina Manna, {\it Int.J.Mod.Phys. D} {\bf 26} (2017) 1750090.

\bibitem{anisotropico6} S. Thirukkanesh, F. C. Ragel, Ranjan Sharma, Shyam Das , {\it Eur. Phys. J. C} {\bf 78} (2018) 31.

\bibitem{anisotropico7} L. Herrera, J. Ospino, A. Di Prisco , {\it Phys.Rev.D} {\bf 77} , (2008) 027502.

\bibitem{anisotropico8} M. Chaisi, S. D. Maharaj, {\it Pramana} {\bf 66}, (2006), 313. 

\bibitem{Ovalle1}  J Ovalle , {\it Phys. Rev. D} {\bf 95}, (2017) 104019 .

\bibitem{Ovalle3} J Ovalle ,{\it  Mod.Phys.Lett.A} {\bf 23} , (2008) 3247.

\bibitem{Ovalle4}  J Ovalle , Braneworld Stars: Anisotropy Minimally Projected Onto the Brane , in {\it Proceedings of the Ninth Asia-Pacific International Conference }, (2009) Singapore, p.~173–182.

\bibitem{Ovalle5} Roberto Casadio, J Ovalle, Roldao da Rocha  , , {\it Class. Quantum Grav.} {\bf 32} (2015) 215020. 

\bibitem{Ovalle6}  J Ovalle, {\it Int. J. Mod. Phys. Conf. Ser.}, {\bf 41}, (2016) 1660132.

\bibitem{Ovalle7} J Ovalle, F Linares, {\it Phys.Rev. D} {\bf 88} (2013) 104026. 

\bibitem{Ovalle8} J Ovalle, Laszló A. Gergely, Roberto Casadio, {\it Class. Quantum Grav.} {\bf 32} (2015) 045015 . 

\bibitem{Ovalle9} Roberto Casadio, J Ovalle, Roldao da Rocha, {\it Europhys. Lett.} {\bf 110} (2015) 40003 

\bibitem{Ovalle10} Gabbanelli, L., Rincon, A. and Rubio, C, {\it Eur. Phys. J. C} {\bf 78} (2018) 370. 

\bibitem{Ovalle2}  J Ovalle, R Casadio, R da Rocha, A Sotomayor , {\it Eur. Phys. J. C} {\bf 78}, (2018) 122. 

\bibitem{Heint} Heintzmann, H. Z {\it Physik} {\bf 228} (1969)  489.

\bibitem{Lake}   Delgaty, Kayll Lake , { \it Comput.Phys.Commun.}, (1998), {\bf 115}, 395. 


\bibitem{Heint1} Pant, N., Mehta, R.N. ans Pant, M.  , { \it  Astrophys Space Sci}, (2011), {\bf 332}, 473. 

\bibitem{Heint2} Singh, K.N. and Pant, N.  , { \it  Indian J Phys}, (2016), {\bf 90}, 843. 

\bibitem{Israel}   Cristiano Germani, Roy Maartens , { \it Phys.Rev. D}, (2001), {\bf 64}, 124010. 


\bibitem{Herrera1}  Herrera, N.O. Santos ,{\it Phys.Rept.} {\bf 286} (1997) 53.

\bibitem{Maartens} George Ellis, Roy Maartens, Malcolm MacCallum  , {\it Gen.Rel.Grav.} {\bf 39}, (2007), 1651.

\bibitem{Abreu} H. Abreu, H. Hernandez, L.A. Nunez  , {\it Class. Quantum Grav. } {\bf 24}, (2007), 4631.

\bibitem{herrera5}  L. Herrera, {\it Phys. Lett. A} {\bf165}, 206 (1992).

\bibitem{Herrera3} R. Chan,  L. Herrera,  N. O. Santos , {\it Mon. Not. R. Astron. Soc.} {\bf 265}, (1993), 533-544.

\bibitem{Herrera4} R. Chan, L. Herrera, N. O. Santos. , {\it Mon. Not. R. Astron. Soc.} {\bf 267}, (1994), 637-646.

\bibitem{hillebrandt} H. Heintzmann, W. Hillebrandt, {\it Astron. Astrophys.} {\bf38}, 51 (1975).

\bibitem{chan} R. Chan, S. Kichenassamy, G. Le Denmat, N.O. Santos, {\it Mon. Not. R. Astron. Soc.} {\bf239}, 91
(1989).

\bibitem{Anisotropico9} S.K. Maurya, Y.K. Gupta, Saibal Ray, Baiju Dayanandan , {\it Eur. Phys. J. C} {\bf 75}, (2015), 225.

\bibitem{Herrera2}  Conversation via e-mail with Luis Herrera

\bibitem{Harko3}  Conversation via e-mail with Tiberiu Harko.



\end{thebibliography}
\end{document}